\documentclass[aps,prc,twocolumn,groupedaddress]{revtex4}
\newcommand \bea{\begin{eqnarray}}
\newcommand \eea{\end{eqnarray}}
\newcommand \beq{\begin{eqnarray}}
\newcommand \eeq{\end{eqnarray}}

\newcommand{\ve}[1]{\mbox{\boldmath $#1$}}

\begin{document}


\title{Static response of Fermi liquids with tensor interactions}



\author{E. Olsson}
\affiliation{Department of Astronomy and Space Physics, Uppsala University,
 Box 515,
75120 Uppsala, Sweden \\
and NORDITA, Blegdamsvej 17, DK-2100 Copenhagen \O, Denmark}
\author{P. Haensel}
\affiliation{Nicolaus Copernicus Astronomical Center, Bartycka 18, 00-716 Warsaw,
Poland}

\author{C. J. Pethick}
\affiliation{NORDITA, Blegdamsvej 17, DK-2100 Copenhagen \O, Denmark}

\date{\today}

\begin{abstract}
We use Landau's theory of a normal Fermi liquid to derive expressions for the static response of a system with a general tensor interaction that conserves the total spin and the total angular momentum of the quasiparticle-quasihole pair.  The magnetic susceptibility is calculated in detail, with the 
inclusion of the center of mass tensor and cross vector terms in addition to the exchange tensor one.  We also introduce a new parametrization of the tensor Landau parameters which significantly reduces the importance of high angular harmonic contributions.  For nuclear matter and neutron matter we find that the two most important effects of the tensor interaction are to give a contribution from multipair states and to renormalize the magnetic moments. Response to a weak probe may be calculated using similar methods, replacing the magnetic moments with the matrix elements of the weak charges.    
\end{abstract}


\maketitle


\section{Introduction}
For systems with central interactions, Landau's theory of normal Fermi
liquids provides an economical way of characterizing many low-temperature
properties.  The theory was applied to atomic nuclei by Migdal and
collaborators, and in that work it was generally assumed that the non-central
contributions to the effective nucleon-nucleon interaction were small
\cite{migdal}.  The generalization of Landau theory to include effects of the
tensor force was made by D\c{a}browski and Haensel
\cite{dabrowski2,dabrowski1,haensel2}.  Subsequently,
estimates of tensor contributions to the effective interaction were made in 
Refs.\ \cite{baeckman,schwenk,schwenk2,dickhoff}. 

The stimulus for the present work arose in the context of astrophysics.  
In the physics of collapse and the subsequent evolution of a
neutron star, the properties of neutrinos in dense matter are a key ingredient
\cite{janka}. As demonstrated in Refs.\
\cite{sawyer,iwamoto,raffelt1,raffelt2,raffelt3,burrows,prakash,prakashreddy},
neutrino scattering and absorption rates     
are sensitive to nucleon-nucleon interactions, especially their spin-dependent
parts. 
Direct calculation of effective interactions is difficult, but for systems with
only central interactions, information about the effective interaction at long
wavelengths may be obtained directly from a knowledge of static properties of the
matter. 
For example, the spin susceptibility of a single-component Fermi liquid with two
spin states is given by
\beq
\chi=\frac{\mu_0^2 N(0)}{1+G_0},
\label{Landauchi}
\eeq
where $\mu_0$ is the magnetic moment of a particle in free space, $N(0)=m^*
p_{\rm F}/\pi^2\hbar^3$ is the density of states per unit volume at the Fermi
surface, $m^*$ is the quasiparticle effective mass, $p_{\rm F}$ is the Fermi
momentum and $G_0$ is the Landau parameter describing the isotropic part of the
spin dependent contribution to the quasiparticle interaction.  Since calculations
of the magnetic susceptibility of neutron matter exist \cite{Fantoni}, it is
relevant to ask to
what extent it is possible to deduce properties of the effective interaction from
such data. 
A related question is the extent to which tensor contributions to quasiparticle energies and interactions alter neutrino scattering rates, which were previously calculated neglecting tensor effects in Ref. \cite{iwamoto}.

As shown in earlier work \cite{olsson}, the tensor force influences the static response of a system in a number of different ways. One is that the magnetic moment of a quasiparticle is different from its value for a particle in vacuum. In particular, the magnetic moment is not a scalar, as it is for systems with central forces only. As long ago as 1951, Miyazawa \cite{miyazawa} calculated explicitly how the magnetic moment of a nucleon would be modified by the tensor interaction due to one pion exchange and more recent discussions may be found in e.g. the review \cite{arima} and Ref. \cite{riska}. A second effect is that the quasiparticle interaction contains explicit tensor contributions. Following Refs. \cite{dabrowski1,dabrowski2,haensel2}, it has generally been assumed that these have an exchange-tensor structure similar to that of the one-pion exchange interaction.
However, Schwenk and Friman \cite{schwenk2} have pointed out recently that the one-pion exchange interaction, when acting in second order, can give rise to contributions to the effective interaction which have a different structure. In their paper they evaluated these induced interaction contributions to the quasiparticle interaction and found that the exchange-tensor term in the quasiparticle interaction is much reduced, and that terms of a different structure can be of a comparable magnitude to the exchange-tensor ones. A third effect of the tensor interaction is that there are multipair contributions to the magnetic susceptibility.

In this paper we begin by deriving a general expression for the magnetic susceptibility in Sec.\ \ref{Sec:response}, taking into account tensor contributions to the magnetic moment, and tensor contributions to the effective interaction which are completely general for an interaction which conserves the total angular momentum and the total spin of the quasiparticle-quasihole pair. This represents a generalization of the earlier calculation of the magnetic susceptibility by Haensel and Jerzak \cite{haensel}, who took into account the effect of the exchange-tensor contribution to the quasiparticle interaction. Another issue that we address is how to parametrize tensor contributions to the quasiparticle interaction (Sec.\ \ref{Sec:tensor}). The scheme usually employed in the past suffers from the disadvantage that it is generally necessary to take into account high angular harmonic contributions to the interaction. We present an alternative parametrization for which higher harmonic terms play little role. 
In Sec.\ \ref{Sec:ME} we evaluate the matrix elements needed for the calculation of the susceptibility and estimate the magnitudes of the different contributions. In Sec.\ \ref{Sec:SNM} we extend the result to symmetric nuclear matter and to responses with other spin and isospin properties, for example that to weak interactions.   Sec.\ \ref{Sec:conclusions} contains concluding remarks.

\section{Basic formalism}
\label{Sec:basic}
The reason for Landau's theory of normal Fermi liquids being particularly 
simple 
for systems with interparticle interactions that are central is that 
 the energies of long-wavelength, low-lying states differing little from the ground state may
be described solely in terms of the quasiparticle distribution. The importance of conservation laws in determining the low-frequency, long-wavelength behaviour of Fermi systems has been stressed by Leggett \cite{leggett65b} and a discussion in terms of Landau theory is given in Ref.\ \cite{baym}. Expressed in other
terms, the only excitations of importance are ones with a single added 
quasiparticle and a single quasihole. When non-central interactions are present,
the energy must include contributions from states with more than one
quasiparticle-quasihole pair. To facilitate satisfying the conservation law for particle number, it is convenient to work with the thermodynamic potential $\langle\hat{H}-\tilde{\mu}\hat{N}\rangle$ which, for brevity, we shall refer to simply as the energy. Here $\hat{H}$ is the Hamiltonian operator, $\hat{N}$ the particle number operator and $\tilde{\mu}$ the chemical potential.  We write the change $\delta E$ in the energy when the system is excited from its ground state as the sum of a quasiparticle (Landau) contribution, $\delta E_{\rm L} $, and a multipair contribution, $\delta E_{\rm M}$: 
\beq
\delta E =\delta E_{\rm L} +\delta E_{\rm M}.
\eeq

The quasiparticle contribution to the energy may be expressed as the sum of two terms, the intrinsic kinetic and mutual  interaction energies of the quasiparticles, and the energy of interaction with the magnetic field:
\beq
\label{E_L}
\delta E_{\rm L} =\delta E_{\rm L}^{\rm int} +\delta E_{\rm L}^{\mathcal H }.
\eeq
For simplicity, let us begin by considering a system of one species of fermions
with spin 
1/2.
The change in the kinetic and interaction energies of the system when the 
quasiparticle distribution $n_{{\bf p}\ve{\scriptstyle\sigma}}$ changes by an
amount $\delta n_{{\bf p}\sigma}$ is given to second order by the standard
expression \cite{landau1,landau2}, which amounts to the statement that the quasiparticle contribution to the energy is given by
\bea
\nonumber
\delta E_{\rm L}^{\rm int} ={\rm Tr}_{\sigma} \int 
\frac{d^3 p}{(2\pi\hbar)^3}(\epsilon_{\bf p}^0-\tilde{\mu})
\delta n_{\bf p}(\ve{\sigma})\\
\nonumber
+\frac{1}{2}{\rm Tr}_{\sigma}{\rm Tr}_{\sigma'}\int \frac{d^3
p}{(2\pi\hbar)^3}\int \frac{d^3 p'}{(2\pi\hbar)^3}f_{{\bf p
p'}\ve{\scriptstyle\sigma\sigma}'}\delta n_{\bf p}(\ve{\sigma})\delta n_{\bf
p'}({\ve{\sigma}}'),\\
\label{ELint}
\eea
where $\epsilon_{\bf p}$ is the quasiparticle energy. Note that $\delta n_{\bf p}(\ve{\sigma})$ is a matrix in spin space.
The quantity $f_{\bf pp'{\ve{\scriptstyle \sigma\sigma}}'}$ is the Landau quasiparticle interaction, which we write in the form
\beq
f_{\bf p p'\ve{\scriptstyle\sigma \sigma}'}= f_{\bf p p'}+ g_{\bf p
p'}\ve{\sigma}\cdot\ve{\sigma}' + f_{\bf p p'\ve{\scriptstyle\sigma
\sigma}'}^{\rm T},
\label{landauint}
\eeq
where $f_{\bf p p'}$ is the spin-averaged quasiparticle interaction, $g_{\bf p
p'}$ is the spin-exchange contribution and $f_{\bf p p'\ve{\scriptstyle\sigma
\sigma}'}^{\rm T}$ is the tensor or, more generally, the non-central contribution to the interaction.
We shall discuss the form of the tensor interaction, 
$f_{\bf p p'\ve{\scriptstyle \sigma \sigma}'}^{\rm T}$, in Sec.\ \ref{Sec:tensor} in the light of the new results of Schwenk and Friman 
\cite{schwenk2}. 
To begin, we derive an expression for the magnetic susceptibility, for a comletely general interaction that conserves the total spin and the total angular momentum of the quasiparticle-quasihole pair, and subsequently we shall discuss its specific form in detail.

Non-central forces can alter the effective charges of the quasiparticles.
When calculating the energy of interaction of the quasiparticles with an external field it is therefore important to allow for the fact that the effective couplings (charges) of quasiparticles may have components which are not scalars under rotations of the momentum of the quasiparticle. For definiteness, we consider the case of response to a magnetic field.
The change in quasiparticle energy due to the application of a magnetic field,
${\mathcal H}$, is given by
\beq
\delta E_{\rm L}^{\mathcal H}=-{\rm Tr}_{\sigma} \sum_{ij} \int \mu_{ij}\sigma_j {\mathcal H}_i
\delta n_{\bf p}(\ve{\sigma}) \frac{d^3 p}{(2\pi\hbar)^3},
\eeq
where $\mu_{ij}$, the magnetic moment matrix, is
\beq
\mu_{ij}= \mu \delta_{ij} + \frac{3}{2}\mu_{\rm T}\left(\frac{p_ip_j}{p^2}-\frac{\delta_{ij}}{3}\right)
\eeq
where $\mu$ and $\mu_{\rm T}$ are parameters. Let us now consider the change in the energy when the Fermi surface is distorted.  The distortion is specified by the function
\beq
\nu(\hat{\bf p})_{\alpha\beta} =\bf{u}(\hat{\bf p})\cdot \ve{\sigma}_{\alpha\beta}
\eeq
which corresponds to the change in the Fermi momentum as a function  of direction.  At zero temperature,  the corresponding change in the distribution function $\delta n_{\bf p}$ may be expanded in powers of $\nu$ 
and to second order one has
\bea
\nonumber
(\delta n_{\bf p})_{\alpha\beta}&=& \delta (p-p_{\rm F})\nu(\hat{\bf
p})_{\alpha\beta} \\
&-&\frac{1}{2} \delta'(p-p_{\rm F})\sum_\gamma \nu(\hat{\bf
p})_{\alpha\gamma}\nu(\hat{\bf p})_{\gamma\beta}.
\eea
It is advantageous to use spherical tensors so, following Refs. \cite{baeckman,haensel}, we write 
\beq
\nu(\hat{\bf p})=\sum_{\nu = -1}^{1}(-)^{\nu}u^{\nu}(\hat{\bf p})\sigma^{-\nu}
\eeq
and expand $u$ in spherical harmonics
\beq
u^{\nu}=\sum_{l,m} u^{\nu}_{lm} Y_{lm}(\hat{\bf p}).
\eeq

We shall work with eigenstates of the total angular momentum of the quasiparticle-quasihole pair.
The corresponding amplitudes are constructed by the transformation
\beq
C_{lJ}^M = \sum_{m \mu} (-)^{\mu}(l m 1 -\mu | JM) u_{lm}^{\mu},
\eeq
where $(l m 1 -\mu| JM)$ are Clebsch-Gordan coefficients.

The part of the quasiparticle energy not involving the external field, Eq. (\ref{ELint}), has been calculated in Refs. \cite{baeckman,haensel} to be
\beq
\delta E_{\rm L}^{\rm int}=\frac{p_{\rm F}^3}{(2\pi\hbar)^3 m^*} \sum_{l l' JM} C_{lJ}^M C_{l'J}^{M*}\langle l'J|A|lJ\rangle, 
\eeq
where the matrix elements in the new basis are given by
\bea 
\nonumber
\langle l'J'|A|lJ \rangle &=& \sum_{m,\mu,m',\mu'} (J'M'|l' m' 1
-\mu' )\\
&&\langle l'm'\mu'|A|l m \mu \rangle (l m 1 -\mu | JM),
\eea
$\langle l'J'|A|lJ \rangle$ are the normalized matrix elements, and 
\beq 
\label{A}
A=1+\mathcal F,
\eeq
the unit coming from the first term in Eq.\ (\ref{E_L}), and $\mathcal{F} = N(0)f_{\bf p p'\ve{\scriptstyle\sigma \sigma}'}$ comes from the interaction term.
For a quasiparticle interaction containing only exchange-tensor terms, the matrix elements $\langle lJ|A|l'J\rangle$ have been calculated for the
conventional parametrization for the exchange tensor interaction in Ref. \cite{baeckman}.

If we take the magnetic field to be in the $z$-direction, ${\mathcal H}={\mathcal H}_z$, we find that the energy due to the application of this external field is
\bea
\nonumber
\delta E_{\rm L}^{\mathcal H}&=&  \frac{2p_{\rm F}^2\sqrt{4\pi}}{(2\pi\hbar)^3}\mathcal{H}_z\times\\
&&\!\!\!\left[\mu u^{0}_{00}+\frac{\mu_{\rm T}}{\sqrt{5}} u^{0}_{20}- \frac{3}{2}\frac{\mu_{\rm T}}{\sqrt{15}}\left(u^{-1}_{21}+ u^{1}_{2-1}\right) \right]\!.
\eea
Rewriting this in terms of the coefficients $C_{lJ}^M$ and adding the intrinsic contribution to the energy leads to the following expression for the quasiparticle energy: 
\bea
\nonumber
\delta E_{\rm L} =\frac{p_{\rm F}^3}{(2\pi\hbar)^3 m^*} \left[\sum_{l l' JM} C_{lJ}^M C_{l'J}^{M*}
\langle l'J|A|lJ\right.
\rangle \\ \nonumber
\left.- \frac{2 m^*}{p_{\rm F}} \sqrt{4\pi} \mathcal{H}_{z}
\left(\mu  C_{01}^0 - \frac{\mu_{\rm T}}{\sqrt{2}}C_{21}^0\right)\right].
\eea
At first sight, one may be surprised that there are no terms with
$J=3$.  However, this follows immediately from the observation that the
magnetic moment operator $\mu_{ij}\sigma_j$ transforms as an axial vector
under simultaneous rotations in momentum space and spin space, and
consequently it contains only components corresponding to total angular
momentum $J=1$.

By minimizing the quasiparticle contribution to the energy with respect to the coefficients $C_{lJ}^M$ we find that
all $C_{lJ}^M$ are zero except the following two,
\bea
C_{01}^0&=&\sqrt{4\pi} \mathcal{H}_z \frac{m^*}{p_{\rm F}}\left(\mu \langle 21|A|21\rangle + \frac{\mu_{\rm
T}}{\sqrt{2}} \langle 01|A|21\rangle\right)\nonumber \\
&\times &\frac{1}{\langle 21|A|21\rangle \langle 01|A|01\rangle-\left|\langle 01|A|21\rangle\right|^2}
\label{c01}
\eea
and
\bea
C_{21}^0&=&\sqrt{4\pi} \mathcal{H}_z \frac{m^*}{p_{\rm F}}\left(-\mu \langle 01|A|21\rangle - \frac{\mu_{\rm T}}{\sqrt{2}} \langle 01|A|01\rangle\right)\nonumber\\
&\times &\frac{1}{\langle 21|A|21\rangle \langle 01|A|01\rangle-\left|\langle 01|A|21\rangle\right|^2}. \label{c21}
\eea

\section{Static response functions}
\label{Sec:response}
The magnetic susceptibility is defined as the derivative of the magnetization
with respect to the magnetic field at zero field,
\beq
\chi = \left. \frac{\partial M_i}{\partial{\mathcal H}_i} \right|_{{\mathcal H}=0}.
\eeq
We write the magnetization, $M$, as the sum of two terms, one  contribution from single quasiparticle-quasihole pairs, $M_{\rm L}$, and another contribution from the excitations of multipair states, which we denote by $M_{\rm M}$:
\beq
M =M_{\rm L}+ M_{\rm M}.
\eeq
The first term, the Landau term, may be calculated by the same methods as we used in the previous section, and in terms of the $C_{lJ}^{M}$ it is given by:
\bea
M_{\rm L}&=& \int \sum_{j}\frac{d^3p}{(2\pi\hbar)^3} \mu_{ij}\sigma_j \delta n_{\bf p}(\ve{\sigma})\nonumber\\
&=&\frac{2 p_{\rm F}^2\sqrt{4\pi}}{(2\pi\hbar)^3} 
\left[\mu  C_{01}^0 - \frac{\mu_{\rm T}}{\sqrt{2}}C_{21}^0 \right].
\eea
By using Eqs.\ (\ref{c01}) and (\ref{c21}) for the $C_{lJ}^M$, and the relation between the susceptibility and the magnetization,
we obtain the following expression for the susceptibility:
\bea\label{chi}
\chi&=& \nonumber\\
&&N(0)\left[\mu^2 \frac{\langle 21|A|21\rangle }{\langle 21|A|21\rangle \langle01|A|01\rangle -\langle 01|A|21\rangle^2}\nonumber\right. \\
&+&\!\!\left.\mu \mu_{\rm T } \sqrt{2} \frac{\langle 01|A|21\rangle }{\langle
21|A|21\rangle \langle 01|A|01\rangle -\langle 01|A|21\rangle^2}\right.\nonumber\\
&+&\!\!\left.\frac{\mu_{\rm T }^2}{2}\frac{\langle01|A|01\rangle}{\langle 21|A|21\rangle
\langle 01|A|01\rangle -\langle 01|A|21\rangle^2}\right]  \nonumber\\
&+&\chi_{\rm M},
\eea
where $\chi_{\rm M}$ is the contribution from transitions to multipair states.
Here we have implicitly made the usual choice of phase for the states, in which case the matrix elements of $A$ are real.
If we take only the first term and set $\mu = \mu_0$ in Eq. (\ref{chi}), this result reduces to that of Ref. \cite{haensel}. 
Quantitative estimates of the different effects will be
described in Sec.\ \ref{Sec:ME} after calculation of the relevant matrix elements, but first we shall discuss the form of the tensor interaction.

\section{The tensor interaction}
\label{Sec:tensor}
The explicit form of the tensor interaction including all terms allowed by invariance under simultaneous rotations in coordinate space and spin space is \cite{schwenk2} 
\beq
f^{\rm T}_{\bf p p'\ve{\scriptstyle{\sigma \sigma}}'}= h_{\bf p p'} S(\hat{\bf{q}})+k_{\bf p p'} S(\hat{\bf{P}})+l_{\bf p p'} A(\hat{\bf q},\hat{\bf P}).
\label{Eq:fulltensor}
\eeq
Here
\beq
S(\hat{\bf{q}})=3\ve{\sigma}\cdot\hat{\bf q}\ \ve{\sigma}'\cdot \hat{\bf q}-\ve{\sigma}\cdot\ve{\sigma}'
\label{tensorop}
\eeq
is referred to as the exchange-tensor operator, ${\bf q}={\bf p}-{\bf p'}$ being the momentum transfer. The operator in the second term, which is referred to as the center of mass (cm) tensor, is given by
\beq
S(\hat{\bf{P}}) =  3\ve{\sigma}\cdot\hat{\bf P}\ \ve{\sigma}'\cdot \hat{\bf P}-\ve{\sigma}\cdot\ve{\sigma}'.
\eeq
It has the same form as the exchange-tensor, but with ${\bf q}$ replaced by the total momentum $\bf{P}=\bf{p}+\bf{p'}$. The final term, referred to as the cross-vector term, is given by
\bea
A(\hat{\bf q},\hat{\bf P})&=&(\ve{\sigma}\times\ve{\sigma}')\cdot(\hat{\bf{q}}\times\hat{\bf{P}})\nonumber\\
&=&\ve{\sigma}\cdot\hat{\bf P}\ \ve{\sigma}'\cdot\hat{\bf q}-\ve{\sigma}\cdot\hat{\bf q}\ \ve{\sigma}'\cdot\hat{\bf P}.
\eea
The functions $h_{\bf p p'}$, $k_{\bf p p'}$ and $l_{\bf p p'}$ may be expanded in Legendre polynomials of $\cos\theta$, where $\theta$ is the angle between $\bf p$ and $\bf p'$, e.g. for $h_{\bf p p'}$,
\beq
h_{{\bf p} {\bf p'}}=\sum_{l=0}^{\infty}h_l P_l (\cos \theta).
\label{Legendreexp}
\eeq
Note that the choice of parametrization of the exchange-tensor term in Eq. (\ref{Eq:fulltensor}) is different than the one traditionally used. The conventional way to parametrize this term is to write \cite{dabrowski2} 
\beq
h_{\bf p p'} S(\hat{\bf{q}})=\frac{q^2}{p_{\rm F}^2} \tilde{h}_{\bf p p'} S(\hat{\bf{q}}).
\label{tensorh}
\eeq
The functions $h$ and $\tilde{h}$ contain the same physical information, but have different properties as far as their expansions in terms of Legendre polynomials are concerned. We will now compare some aspects of the different choices of parametrization.

With the conventional parametrization, $\tilde{h}_{\bf p p'}$, the explicit $q^2$ factor ensures that the tensor interaction vanishes identically for ${\bf q} \rightarrow 0$, provided that $\tilde{h}_{\bf p p'}$ is finite for ${\bf p} \rightarrow {\bf p}'$.  This is in agreement with the form of the one-pion-exchange contribution which is given by
\beq
f^{\rm T}_{\bf p p'\ve{\scriptstyle{\sigma \sigma}}'} = \frac{f^2}{3m_{\pi}^2} \frac{q^2}{q^2+m_{\pi}^2} S(\hat{\bf{q}}),
\label{Eq:OPEP}
\eeq 
where $f$ is the pion-nucleon coupling constant, and $f^2/(4\pi\hbar c) \approx 0.08$.
More generally, if the tensor interaction is analytic for small ${\bf q}$, one would expect $h_{\bf p p'}$ to tend to zero for ${\bf p} \rightarrow {\bf p}'$, since otherwise the tensor interaction would depend on the way in which ${\bf p}$ approaches ${\bf p}'$. Such nonanalytic contributions to the quasiparticle interaction do occur for systems with central interactions \cite{carneiro1,carneiro2}, and there they are due to exchange of quasiparticle-quasihole pairs and collective modes with long wavelengths. The work of Schwenk and Friman \cite{schwenk2} indicates that such contributions do not occur in $h_{\bf p p'}$, although they do in $k_{\bf p p'}$ (see Sec. \ref{Sec:K}). To ensure that $h_{\bf p p'}\rightarrow 0$ for ${\bf p} \rightarrow {\bf p}'$, the $h_l$ must satisfy the condition
\beq
\sum_{l=0}^{\infty}h_l =0. 
\label{Eq:sumrule}
\eeq   
If the series is truncated after a finite number of terms, this will inevitably lead to the sum rule being violated. However, this should not be very important as long as long-wavelength processes do not play a dominant role. 

The conventional parametrization suffers from the disadvantage that, when expanding $\tilde{h}$ in terms of Legendre polynomials (\ref{Legendreexp}), it is generally necessary to include many terms in the sum in order to obtain a good representation of the interaction. Consider, for example, the one pion exchange contribution. Using the fact that $\cos \theta =1-q^2/(2p_{\rm F}^2)$ one finds
\bea
\tilde{h}_l^{\rm OPE}&=&\nonumber\\
&=&(2l+1)\!\!\frac{f^2}{(3m_{\pi}^2)}\int_0^{2p_{\rm F}}\!\!\!\frac{dq^2}{p_{\rm F}^2}
\frac{1}{q^2+m_{\pi}^2} P_l [1-q^2/(2p_{\rm F}^2)]\nonumber\\
&=&(2l+1)\frac{f^2}{(3m_{\pi}^2)} Q_l[1+\frac{m_{\pi}^2}{2p_{\rm F}^2}],
\eea
where $Q_l$ is a Legendre function of the second kind. 
\begin{table}
\caption{Tensor parameters for neutron matter with Fermi wave vector $k_{\rm F}=1.7\ \rm{F}^{-1}$. Values for the tensor parameters by Schwenk \cite{schwenk3}\label{AchimsH}, obtained taking into account the tensor interaction to first order by the renormalization group method described in Refs. \cite{schwenk,schwenk1}, are presented in the first and third columns, for the two choices of parametrization. For comparison we present the result for the one-pion exchange potential (Eq. (\ref{Eq:KOPE})) in column two and four. All tensor parameters are calculated for $m^*/m =0.8345$.}
\begin{ruledtabular}
\begin{tabular}{lllll}
l  & $H_l$ & $H_l^{\rm{OPE}}$ & $\tilde{H}_l$ & $\tilde{H}_l^{\rm{OPE}}$\\
\hline
0&0.529& 0.403&0.665&0.751\\
1&0.150&  -0.088&1.200&1.037\\
2&-0.0959& -0.079&1.150&0.935\\
3&-0.141& -0.064&0.933&0.753\\
4&-0.124& -0.049&0.703&0.575\\
5&-0.0944& -0.036&0.509&0.427\\
6&-0.0697& -0.026&0.359&0.310\\
7&-0.0494& -0.019&0.249&0.223\\
8&-0.0335& -0.013&0.171&0.158\\
9&-0.0225& -0.009&0.116&0.111\\
10&-0.0151& -0.007&0.0774&0.078\\
\end{tabular}
\end{ruledtabular}
\end{table}
In Table 1 we show
results of calculations of the exchange tensor interaction, $\tilde{H}_l=N(0)\tilde{h}_l$, by Schwenk \cite{schwenk3}, for $k_{\rm F} = 1.7$ F$^{-1}$.  Note that these were calculated taking the tensor interaction into account only in first order. 
One sees that as $l$ increases from zero, $\tilde{H}_l$ first increases and then decreases. It is therefore a poor approximation to truncate
the expansion after the first few terms.  The reason for this is that typical
Fermi wave numbers at nuclear densities are of order $1.5$ F$^{-1}$, which is large
compared with the pion mass, $m_{\pi}/\hbar c$ $\sim 0.7$ F$^{-1}$.  
In the limit of the  zero pion mass, all the  $\tilde{h}_l$ diverge, the leading contribution being
$(2l+1)f^2/(3m_{\pi}^2)\ln(2p_{\rm F}/m_{\pi})$. This is due to the fact that $\tilde{h} \sim q^{-2}$,
which leads to a logarithmic integral for $\tilde{h}_l$. 

The situation is quite different for the parametrization (\ref{Legendreexp}). For the one pion exchange potential (\ref{Eq:OPEP}) one finds
\beq
\label{Eq:KOPE}
h_l=\frac{f^2}{(3m_{\pi}^2)}\left(\delta_{l,0}-\frac{m_{\pi}^2}{2p_{\rm F}^2} (2l+1) Q_l[1+{m_{\pi}^2}/{2p_{\rm F}^2}]    \right).
\eeq 
This shows that for small $m_{\pi}/p_{\rm F}$, the leading term is 
\beq
h_l \sim \frac{f^2}{3m_{\pi}^2}\frac{2l+1}{2} \left(\frac{m_{\pi}}{p_{\rm F}}\right)^2 \ln(2p_{\rm
F}/m_{\pi}).
\eeq
In the limit of zero pion mass, $h_{{\bf p}{\bf p'}}$ becomes a constant, and therefore the only nonvanishing coefficient in
the expansion in Legendre polynomial is $h_0$. Note that for both $\tilde{h}_l$ and $h_l$,  the limits $m_{\pi}/p_{\rm F}\rightarrow 0$ and
$l\rightarrow \infty$ do not commute.

Let us now compare the expansions of $h_l$ and $\tilde{h}_l$.
Using the relation between $h_{{\bf p} {\bf p'}}$ and $\tilde{h}_{{\bf p} {\bf p'}}$, Eq. (\ref{tensorh}) and that $1-\cos \theta=q^2/(2p_{\rm F}^2)$,
we find the following expression for $h_l$:
\beq
h_l\!=\!2(2l+1)\!\!\int_{-1}^{+1}\!\!\!\frac{d(\cos\theta)}{2}(1-\cos\theta )\tilde{h}(\cos\theta)P_l (\cos\theta),
\eeq
or
\beq
h_l=2\left( \tilde{h}_l-\frac{l+1}{2l+3}\tilde{h}_{l+1}-\frac{l}{2l-1}\tilde{h}_{l-1}\right).
\eeq
Thus, given the $\tilde{h}_l$'s, we can obtain the $h_l$'s. Next we show how to write the $\tilde{h}_l$'s, given by
\beq
\label{hltildeofh}
\tilde{h}_l=\frac{(2l+1)}{2}\int_{-1}^{1} \frac{d(\cos\theta)}{2}\frac{h(\cos\theta)}{1-\cos\theta} P_l(\cos\theta)
\eeq
in terms of the $h_l$'s.
Expanding the function $h(\cos\theta)$ in Legendre polynomials, we get
\bea
\tilde{h}_l=\frac{(2l+1)}{2}\!\!\int_{-1}^{1}\!\! \frac{d(\cos\theta)}{2}\sum_{l'} h_{l'} P_{l'}(\cos\theta) \frac{P_l(\cos\theta)}{1-\cos\theta}.\nonumber\\
\eea
Since the series on the right hand side is not uniformly convergent for $\cos \theta$ in the closed interval $[-1,1]$, we cannot invert the order of the sum and the integral. However, because of the sum rule (\ref{Eq:sumrule}), $h(\theta\!=\!0)$ vanishes, there is no divergence, and we can rewrite Eq. (\ref{hltildeofh}) as
\bea
\tilde{h}_l=\frac{(2l+1)}{2}\!\!\int_{-1}^{1}\!\! \frac{d(\cos\theta)}{2}\frac{(h(\cos\theta)-h(\theta\!=\!0))}{1-\cos\theta} P_l(\cos\theta).\nonumber\\
\eea
Making use of the sum rule, we find that
\bea
\tilde{h}_l=\frac{(2l+1)}{2}\int_{-1}^{1} \frac{d(\cos\theta)}{2}\sum_{l'} h_{l'} [P_{l'}(\cos\theta) - 1]\frac{P_l(\cos\theta)}{1-\cos\theta}.\nonumber\\
\eea
This expression may be evaluated using a number of standard results.
First
\bea
\int_{-1}^{1} \frac{d(\cos\theta)}{2}\frac{[P_{l'}(\cos\theta) - P_0(z)]}{z-\cos\theta}P_l(\cos\theta)=\nonumber\\
P_{l'}(z)Q_{l}(z)-P_0(z)Q_{l}(z), \ l'\leq l.
\eea
The $Q_{l}(z)$ may be expressed as 
\beq
Q_{l}(z)=\frac{1}{2}P_{l}(z)\ln{\frac{1+z}{1-z}}-W_{l-1},
\eeq
where
\bea
W_{l-1}&=&\sum_{k=1}^{l}\frac{1}{k}P_{k-1}(z)P_{l-k}(z)\quad \rm{and}\\
W_{-1}&=&0.
\eea
The Legendre polynomial $P_{l}(z)$ may be written in the following form:
\bea
P_{l}(z)=2^{-l}\sum_{k=1}^{[l/2]}(-1)^l \left(\begin{array}{c} l\\k\end{array}\right)\left(\begin{array}{c} 2l-2k\\n\end{array}\right)z^{l-2k} .
\eea
Letting $z\rightarrow 1$ in the integral, we find that the logarithmic divergence vanishes and the final result is
\beq
\tilde{h}_l=-\frac{(2l+1)}{2}\sum_{l'=l+1}^\infty h_{l'}\sum_{k=l+1}^{l'}\frac{1}{k}.
\eeq

\section{Evaluation of the matrix elements}
\label{Sec:ME}
We return to the calculation of the matrix elements of $A = 1+\mathcal{F}$, Eq.\ (\ref{A}), which we write as
\beq
\langle lJ|A|l'J\rangle=\delta_{ll'}+\langle lJ|\mathcal{F}|l'J\rangle .
\eeq
We further write $\langle lJ|\mathcal{F}|l'J\rangle$ as the sum of a number of contributions:
\bea
\langle lJ|\mathcal{F}|l'J\rangle&=&\langle lJ|\mathcal{F}^{\rm c}|l'J\rangle+\langle lJ|\mathcal{F}^{\rm ex}|l'J\rangle\nonumber\\
&+&\langle lJ|\mathcal{F}^{\rm cm}|l'J\rangle+\langle lJ|\mathcal{F}^{\rm cv}|l'J\rangle .
\eea
Here $\mathcal{F}^{\rm c}$ is the central part, $\mathcal{F}^{\rm ex}$ is the exchange tensor part, $\mathcal{F}^{\rm cm}$ is the center of mass tensor part and $\mathcal{F}^{\rm cv}$ is the cross vector part.

\subsection{Central and exchange tensor contributions}

In Ref. \cite{baeckman} the central and exchange tensor part in the conventional parametrization were calculated to be
\beq
\langle lJ|\mathcal{F}^{\rm c}|l'J\rangle= \delta_{ll'}\frac{G_l}{2l+1},
\eeq
and
\begin{widetext}
\bea
&&\langle lJ|\mathcal{F}^{\rm ex}|l'J\rangle=\delta_{ll'}\left(\tilde{H}_{l-1}\frac{2l}{(2l-1)(2l+1)}+\tilde{H}_{l+1}\frac{2(l+1)}{(2l+3)(2l+1)}\right)\nonumber\\
&+&(-1)^J 15\left(\begin{array}{ccc} 1&1&2\\0&0&0 \end{array}\right)\left(\begin{array}{ccc} l&l'&2\\0&0&0 \end{array}\right)\!\!\left\{\begin{array}{ccc} 1&l'&J\\l&1&2 \end{array}\right\}\left[\tilde{H}_{l}\!\!\left(\frac{2l+1}{2l'+1}\right)^{1/2}\!\!\!+\tilde{H}_{l'}\left(\frac{2l'+1}{2l+1}\right)^{1/2}\!\right]\nonumber\\
&-&\!\!\frac{3[(2l+1)(2l'+1)]^{1/2}}{2J+1}\!\left(\begin{array}{ccc} l&1&J\\0&0&0 \end{array}\right)\left(\begin{array}{ccc} l'&1&J\\0&0&0\end{array}\right)\!\tilde{H}_J-3[(2l+1)(2l'+1)]^{1/2}\sum_{l''}\!\left(\begin{array}{ccc} l&1&l''\\0&0&0 \end{array}\right)\!\!\left(\begin{array}{ccc} l'&1&l''\\0&0&0 \end{array}\right)\!\!\left.\left\{\begin{array}{ccc} 1&l'&J\\1&l&l'' \end{array}\right\}\!\!\tilde{H}_{l''}\!\!\right).\nonumber\\
\eea
\end{widetext}
Here $\left(\begin{array}{ccc} l_1&l_2&l\\m_1&m_2&m \end{array}\right)$ are Wigner $3j$-symbols and $\left\{\begin{array}{ccc} l_1&l_2&l_3\\l'_1&l_2'&l'_3\end{array}\right\}$ are Wigner $6j$-symbols.

We can transform the matrix elements above to the new parametrization. In the expression (\ref{chi}) we can see that the only matrix elements that are necessary for the calculation of the susceptibility are $\langle 01|A|01\rangle$, $\langle 01|A|21\rangle$ and $\langle 21|A|21\rangle$. The contribution to these matrix elements from non-interacting quasiparticles and from the central part of the quasiparticle interaction is given by
\beq
\label{G}
\langle l'J|1+\mathcal{F}^{\rm c}|lJ\rangle\!\!&=&\!\! \delta_{ll'}\left(1+\frac{G_l}{2l+1}\right)
\eeq
and the exchange tensor contributions by \cite{baeckman}
\bea
\label{Htilde}
\nonumber
\langle 01|\mathcal{F}^{\rm ex}|01\rangle\!\!&=&\!\!0,\\
\langle 01|\mathcal{F}^{\rm ex}|21\rangle\!\!&=&\!\!
-\sqrt{2}\left(\tilde{H}_0-\frac{2}{3}\tilde{H}_1+\frac{1}{5}\tilde{H}_2\right)\ \rm{and} \nonumber\\
\langle 21|\mathcal{F}^{\rm ex}|21\rangle\!\!&=&\!\!-\frac{7}{15}\tilde{H}_1+\frac{2}{5}\tilde{H}_2-\frac{3}{35}\tilde{H}_3.
\eea

Let us now express this result in terms of $H_l$. As an example we consider the combination $ \tilde{H}_0-2\tilde{H}_1/3+\tilde{H}_2/5$.  This may be written in the form
\bea
\nonumber
&&\tilde{H}_0-\frac{2}{3}\tilde{H}_1+\frac{1}{5}\tilde{H}_2\!\! \\
\nonumber
&=&\int_{-1}^{+1}\!\!\frac{d(\cos \theta)}{2}\tilde{H}(\cos
\theta)[P_0(\cos \theta)-2P_1(\cos \theta)+P_2(\cos \theta)]\!\!\\
&=&\int_{-1}^{+1}\!\!\frac{d(\cos \theta)}{2}H(\cos \theta)\frac{[P_0(\cos \theta)-2P_1(\cos
\theta)+P_2(\cos \theta)]}{2(1-\cos \theta)}.\!\!\nonumber\\
\label{hhtilde}
\eea
An interesting property of the fraction in Eq.\ (\ref{hhtilde}) is that it is finite for 
$\theta \rightarrow 0$, and therefore the integral converges if $H(\cos \theta)$ is
finite in this limit, even though individual terms diverge.  One finds similar results for the other combinations of the
$\tilde{H}_l$ that occur in Eqs.\ (\ref{Htilde}).
The corresponding result written in
terms of the $H_l$ are
\bea
\nonumber
\langle 01|\mathcal{F}^{\rm ex}|01\rangle\!\!&=&\!\!0,\\
\nonumber
\langle 01|\mathcal{F}^{\rm ex}|21\rangle\!\!&=&\!\!
-\frac{1}{2\sqrt{2}}\left(H_0-H_1\right)\quad \rm{and}\\
\langle 21|\mathcal{F}^{\rm ex}|21\rangle\!\!&=&\!\!
\frac{H_2}{10}-\frac{1}{4}\left(H_0+H_1\right).
\label{matrixH}
\eea
It remains to calculate the center of mass tensor and the cross vector contributions.
 
\subsection{Center of mass tensor}
\label{Sec:K}
The calculation of the center of mass tensor terms follows closely that of the exchange tensor. 
Let us study the operator $\mathcal{F}^{\rm cm}$ given by
\beq
\mathcal{F}^{\rm cm}_{\bf p p'}=N(0)k_{\bf p p'}S(P)=\sum_l K_l P_l(\cos\theta)S(P).
\eeq
If we reverse one of the momenta, we find that
\bea
\mathcal{F}^{\rm cm}_{\bf p -p'}&=&N(0)k_{\bf p -p'}S(q)\nonumber\\
&=&\sum_l (-1)^l K_l P_l(\cos\theta)S(q).
\eea
Thus we have a similar expression to the exchange tensor operator, with a difference of a factor of $(-1)^l$. For the state vectors $|l'J'\rangle $, reversing the momenta, gives
\beq
\langle -\hat{\bf p}'\ve{\sigma}'| l'J'\rangle = (-1)^{l'}\langle\hat{\bf p}'\ve{\sigma}' | l'J'\rangle.
\eeq
The matrix elements of $\mathcal{F}^{\rm cm}$ is thus the same as those for $\mathcal{F}^{\rm ex}$, apart from an overall factor of $(-1)^{l'}$ and that $H_l$ should be replaced by $(-1)^l K_l $.

However, when calculating the matrix elements of the center of mass tensor, it is necessary to be careful for $\theta\rightarrow \pi$, since the calculation of $k_{\bf p p'}$ by Schwenk and Friman \cite{schwenk2} showed that $k_{\bf p p'}$ does not vanish in this limit, but rather tends to a constant when second order tensor contributions are included. However, we shall show, that for the matrix elements we consider, this will not pose a problem.

By writing $k_{\bf p p'}$ as $(P^2/p^2_{\rm F}) \tilde{k}_{\bf p p'}$ we can use the calculation for the exchange tensor given in Ref. \cite{baeckman}, except that for the odd $l'$ states we have to include a factor $(-1)^{l'}$, and $\tilde{H}_l$ should be replaced by $(-1)^l \tilde{K}_l$, with $\tilde{K}_l = N(0) \tilde{k}_l$.
For the particular matrix elements that we are interested in we find that
\bea
\langle 01|\mathcal{F}^{\rm cm}|21\rangle\!\!&=&\!\!
-\sqrt{2}\left(\tilde{K}_0+\frac{2}{3}\tilde{K}_1+\frac{1}{5}\tilde{K}_2\right)\ \rm{and}\nonumber\\
\langle 21|\mathcal{F}^{\rm cm}|21\rangle\!\!&=&\!\!\frac{7}{15}\tilde{K}_1+\frac{2}{5}\tilde{K}_2+\frac{3}{35}\tilde{K}_3.
\label{Ktilde}
\eea
Just as in the example (\ref{hhtilde}), we find that the matrix elements 
above give rise to combinations of Legendre polynomials which contain a 
factor $1+\cos\theta$, so the integral will be finite for $\theta\rightarrow 
\pi$, provided that $\tilde{k}_{\bf p p'}$ is finite in this limit. Using 
the result of Schwenk and Friman \cite{schwenk2} that $k_{\bf p p'}$ goes 
to a constant when $\theta\rightarrow\pi$, we can see that $\tilde{k}_{l}$, 
which is given by

\beq
\tilde{k}_l = \frac{2l+1}{2 p_{\rm F}^2}\int_{-1}^{1} \frac{k_{\bf p p'}}{1+\cos\theta} P_l(\cos\theta) d(\cos\theta),
\eeq
will, in fact, have a logarithmic divergence. To deal with this divergence, we replace the limit $-1$ by $\Lambda$, and write $\tilde{k}_l$ as the sum of a regular part and a part that contains the logarithmic divergence,
\bea
\label{kwlog}
\tilde{k}_l(\Lambda) = \tilde{k}^{\rm reg}_l +(-1)^l (2l+1)\ln(1+\Lambda)C(\Lambda)
\eea
where $C(\Lambda)=k_{\bf p p'}(\Lambda)/(2p_{\rm F}^2)$ tends to a constant when $\Lambda\rightarrow -1$ and the factor $(-1)^l$ comes from $P_l(-1)=(-1)^l$.
By inserting the expression above in the matrix elements of $\mathcal{F}^{\rm cm}$, Eq. (\ref{Ktilde}), we obtain
\bea
\langle 01|\mathcal{F}^{\rm cm}|21\rangle\!\!&=&\!\!-\sqrt{2}\left(\tilde{K}^{\rm reg}_0+ C(\Lambda) \ln(1+\Lambda)\right.\nonumber\\
&+&\frac{2}{3}\tilde{K}^{\rm reg}_1- 2 C(\Lambda)\ln(1+\Lambda))\nonumber\\
&+&\left.\frac{1}{5}\tilde{K}^{\rm reg}_2+ C(\Lambda)\ln(1+\Lambda)\right)\nonumber\\
&=&-\sqrt{2}\left(\tilde{K}^{\rm reg}_0+\frac{2}{3}\tilde{K}^{\rm reg}_1+\frac{1}{5}\tilde{K}^{\rm reg}_2\right)\\
&\rm{and}&\nonumber\\
\langle 21|\mathcal{F}^{\rm cm}|21\rangle\!\!&=&\!\!\frac{7}{15}[\tilde{K}^{\rm reg}_1 -3 C(\Lambda)\ln(1+\Lambda)]\nonumber\\
&+&\frac{2}{5}\tilde{K}^{\rm reg}_2+2 C(\Lambda)\ln(1+\Lambda)]\nonumber\\
&+&\frac{3}{35}[\tilde{K}^{\rm reg}_3- 7 C(\Lambda)\ln(1+\Lambda)]\nonumber\\
&=&\frac{7}{15}\tilde{K}^{\rm reg}_1 +\frac{2}{5}\tilde{K}^{\rm reg}_2+\frac{3}{35}\tilde{K}^{\rm reg}_3.
\eea
This shows that in the combinations of parameters we need, the divergent contribution vanishes and therefore allows us to simply replace $H_l$ by $(-1)^l K_l$ in the expression (\ref{matrixH}) for the $\langle lJ|\mathcal{F}^{\rm ex}|l'J\rangle$ to obtain $\langle lJ|\mathcal{F}^{\rm cm}|l'J\rangle$ in terms of the $K_l$'s:
\bea
\label{cm}
\nonumber
\langle 01|\mathcal{F}^{\rm cm}|01\rangle\!\!&=&\!\!0,\\
\nonumber
\langle 01|\mathcal{F}^{\rm cm}|21\rangle\!\!&=&\!\!
-\frac{1}{2\sqrt{2}}\left(K_0+K_1\right)\quad \rm{and}\\
\langle 21|\mathcal{F}^{\rm cm}|21\rangle\!\!&=&\!\!
\frac{K_2}{10}-\frac{1}{4}\left(K_0-K_1\right).
\label{matrixK}
\eea

\subsection{The cross vector term}
Finally, we turn to the cross vector term, which is proportional to
\bea
A(\bf{\hat{q}},\bf{\hat{P}})&=&(\ve{\sigma}\times\ve{\sigma'})\cdot(\bf{\hat{q}}\times\bf{\hat{P}})\nonumber\\
&=&(\ve{\sigma}\times\ve{\sigma'})\cdot\left(\frac{\bf{\hat{p}}\times\bf{\hat{p}'}}{|\bf{\hat{p}}\times\bf{\hat{p}'}|}\right)\nonumber\\
&=&\frac{\ve{\sigma}\cdot\bf{\hat{p}}\ve{\sigma'}\cdot\bf{\hat{p}'}-\ve{\sigma}\cdot\bf{\hat{p}'}\ve{\sigma'}\cdot\bf{\hat{p}}}{|\bf{\hat{p}}\times\bf{\hat{p}'}|}.
\eea
The calculations of Ref. \cite{schwenk2} indicate that this term will be well behaved in the limits where $\sin\theta \rightarrow 0$, since $L_{\bf p p'}\rightarrow 0$ for both $\theta = 0$ and $\theta = \pi$. This means that there will be sum rules like the one (\ref{Eq:sumrule}) for both $\theta = 0$ and $\theta = \pi$:
\bea
\sum_lL_l P_l(\cos\theta=1)&=&0, \ {\rm and} \nonumber\\
\sum_l  L_l P_l(\cos\theta=-1)&=&\sum_l L_l (-1)^l P_l(\cos\theta=1)=0\nonumber
\eea
which gives two sum rules, one for the odd coefficients and one for even ones: 
\bea
\sum_{l=0}^{\infty}  L_{2l}&=&0, \ {\rm and} \nonumber\\
\sum_{l=0}^{\infty}  L_{2l+1}&=&0.
\eea

To simplify the calculation of the matrix elements, we define $\tilde{L}_{\bf p p'}=L_{\bf p p'}/\sin\theta$, and calculate the matrix element $\langle lJ|\mathcal{F}^{\rm cv}|l'J\rangle$, which later will be transformed back to the original notation. The cross vector contribution to the quasiparticle interaction is
\beq
\Delta \mathcal{F}^{\rm cv} =N(0) \frac{1}{4\pi}\frac{1}{4} \rm{Tr}_{\scriptstyle\ve{\sigma}\ve{\sigma}'} \int\!\!\int d\Omega d\Omega ' \bf{u}\cdot \ve{\sigma}\mathcal{F}^{\rm cv}\bf{u}'\cdot \ve{\sigma}'.
\eeq
We use the notation and method described in Sec.\ \ref{Sec:basic}, which gives for the matrix elements in the $| l m 1 -\mu \rangle$ basis
\begin{widetext}
\bea
\langle l m 1 -\mu | \mathcal{F}^{\rm cv}|l' m' 1 -\mu'\rangle =\sum_{l'',m''}(-)^{m''} \tilde{L}_{l''}&\times&\\
\left[\!\!\left(\begin{array}{ccc} l&1&l''\\0&0&0 \end{array}\right)\!\!\!\left(\begin{array}{ccc} l'&1&l''\\0&0&0 \end{array}\right)\!\!\!
\left(\begin{array}{ccc} l&1&l''\\m&-\mu &m'' \end{array}\right)\!\!\!\left(\begin{array}{ccc} l'&1&l''\\m'&-\mu ' &m'' \end{array}\right) \right.\!\!\!
&-&\!\!\left(\begin{array}{ccc} l&1&l''\\0&0&0 \end{array}\right)\!\!\left(\begin{array}{ccc} l'&1&l''\\0&0&0 \end{array}\right)\!\!\!\left.\left(\begin{array}{ccc} l&1&l''\\m&-\mu ' &m'' \end{array}\right)\!\!\!\left(\begin{array}{ccc} l'&1&l''\\m'&-\mu  &m'' \end{array}\right)\!\!\right].\nonumber
\eea
\end{widetext}
Changing to the $|lJ\rangle$ basis, we find the matrix element to be
\bea
&&\langle lJ|\mathcal{F}^{\rm cv}|l'J\rangle =\nonumber\\
&+&\!\!\frac{[(2l+1)(2l'+1)]^{1/2}}{2J+1}\!\left(\begin{array}{ccc} l&1&J\\0&0&0 \end{array}\right)\!\!\!\!\left(\begin{array}{ccc} l'&1&J\\0&0&0 \end{array}\right)\!\tilde{L}_J\nonumber\\
&-&[(2l+1)(2l'+1)]^{1/2}\nonumber\\
&&\sum_{l''}\!\left(\begin{array}{ccc} l&1&l''\\0&0&0 \end{array}\right)\!\!\left(\begin{array}{ccc} l'&1&l''\\0&0&0 \end{array}\right)\!\!\left\{\begin{array}{ccc} 1&l'&J\\1&l&l'' \end{array}\right\}\!\tilde{L}_{l''}.\nonumber\\
\eea
For $l'=l+2$, the triangle conditions imply that the 3-j symbols are non-zero only if $J=l+1$ and $l''=l+1$. When this happens the two terms are equal in magnutide but of opposite sign. Therefore all $\langle lJ|\mathcal{F}^{\rm cv}|(l+2)J\rangle $ are zero. The only non-zero matrix elements are when $l=l'$. Then $J$ can be either $l+1$ or $l-1$:
\bea
&&\langle lJ|\mathcal{F}^{\rm cv}|lJ\rangle =\nonumber\\
&+&\frac{2(l+1)^2}{(2l+3)^2(2l+2)}\left[\frac{1}{(l+1)(2l+1)-1}\right]\tilde{L}_{l+1}\nonumber\\
&-&\frac{l}{(2l+1)(2l-1)}\tilde{L}_{l-1},\ {\rm for}\ J=l+1,\ {\rm and} \\
&&\langle lJ|\mathcal{F}^{\rm cv}|lJ\rangle =\nonumber\\
&+&\frac{l}{(2l-1)^2}\left[\frac{1}{l(2l+1)-1}\right]\tilde{L}_{l-1}\nonumber\\
&-&\frac{2(l+1)^2}{(2l+1)(2l+2)(2l+3)}\tilde{L}_{l+1},\ {\rm for}\ J=l-1
\eea
The matrix elements we need thus have the following contributions from the cross vector:
\bea
\label{cv1}
\nonumber
\langle01|\mathcal{F}^{\rm cv}|01\rangle&=&0,\\ \nonumber
\langle21|\mathcal{F}^{\rm cv}|01\rangle&=&0 \qquad {\rm and}\\
\langle21|\mathcal{F}^{\rm cv}|21\rangle&=&\frac{2}{10}\tilde{L}_1-\frac{3}{35}\tilde{L}_3.
\eea
The only matrix element we need for the cross vector term is thus when $l=l'=2$ and $J=1$, and we transform this to our original parametrization by writing:
\bea
\langle 21|\mathcal{F}^{\rm cv}|21\rangle&=&\\
 \int_{-1}^{1}d(\cos\theta) \frac{3}{5}\sum_k \frac{L_k P_k(\cos\theta)}{\sin\theta}\!\!\!\!\!\!&&\!\!\!\!\left[P_1(\cos\theta)-P_3(\cos\theta)\right]\nonumber
\label{MEcv}
\eea
which is the integral
\beq
-\frac{3}{2}\int_{-1}^{1} dx \sum_n L_n P_n (x) \sqrt{1-x^2} x = -\frac{3}{2}\sum_n L_n I_n.
\label{Lsum}
\eeq
We calculate the integral $I_n$ to be
\bea
\label{In}
I_n=\left\{\begin{array}{ll}\displaystyle\sum_{k=0}^{[n/2]} \frac{\pi (-1)^k(2n-2k)!(n-2k+1)}{k!(n-k)!(\frac{n-2k+1}{2})!(\frac{n-2k+3}{2})!},&n=2l+1\nonumber\\
&\\
0 &n=2l\quad .\end{array}\right. \nonumber\\
\eea

\subsection{Matrix elements and the susceptibility}
Combining Eqs. (\ref{G}),(\ref{matrixH}),(\ref{cm}),(\ref{MEcv}) and (\ref{Lsum}) above, we find
\bea
\nonumber
\langle 01|A|01\rangle\!\!&=&\!\!1+G_0,\\
\nonumber
\langle 01|A|21\rangle\!\!&=&\!\!-\frac{1}{2\sqrt{2}}\left[H_0-H_1+K_0+K_1\right]\qquad {\rm and} \\
\langle 21|A|21\rangle\!\!&=&\!\!1+\frac{1}{5}\left(G_2+\frac{H_2+K_2}{2}\right)\nonumber\\
-\frac{1}{4}\left(H_0+K_0\right.&+&\left.H_1-K_1\right)-\frac{3}{2}\sum_n L_n I_n,\nonumber\\
\label{matrixelements}
\eea
with $I_n$ given by Eq. (\ref{In}).

\begin{table}
\caption{ Landau parameters for neutron matter calculated by Schwenk \cite{schwenk2}\label{AchimsLandaus}.The Landau parameters are calculated for Fermi wave vector $k_{\rm F}=1.7\ \rm{F}^{-1}$ and the effective mass is $m^*/m =0.8345$.}
\begin{ruledtabular}
\begin{tabular}{lllll}
l & $G_l$ & $H_l$ & $K_l$ & $L_l$\\
\hline
0&0.842&0.070 &-0.258&-0.060 \\
1&0.412&0.163& 0.146&-0.089\\
2&0.219&-0.301& 0.124&-0.034\\
3&0.109&-0.002& 0.048&0.025\\
4&0.051& -0.150&0.015&0.043\\
\end{tabular}
\end{ruledtabular}
\end{table} 

From these results one may calculate the magnetic susceptibility from Eq. (\ref{chi}). The effects of the tensor interaction are generally small, so to 
estimate the various contributions to the susceptibility we expand the susceptibility to second order in the tensor interaction and $\mu_{\rm T}$ and find
\bea
\label{chiexpanded}
\chi&=&N(0)\left[\mu^2 \frac{1}{1+G_0} + \mu^2 \frac{1}{(1+G_0)^2}\frac{\langle 01|A|21\rangle^2}{1+G_2 /5}\right. \nonumber\\
&+&\mu \mu_{\rm T } \sqrt{2}\frac{\langle 01|A|21\rangle }{(1+G_0)(1+G_2 /5)}\nonumber\\
&+&\left.\frac{\mu_{\rm T }^2}{2}\frac{1}{1+G_2 /5}\right] +\chi_{\rm M}.
\eea 

It is not easy to obtain a reliable estimate of the various terms, since different calculations of the tensor Landau parameters give different result. 
To exemplify this for pure neutrons, we first use the Landau parameters obtained by Schwenk \cite{schwenk3} listed Table \ref{AchimsH}. These take into account the tensor interaction only to first order, and consequently the only tensor term is the exchange tensor. We find
\bea
\nonumber
\langle 01|A|21\rangle &\approx&-0.13\quad {\rm and}\\
\langle 21|A|21\rangle &\approx&0.86 .
\eea
We can now compare this with what we get if we use the Landau parameters of Ref. \cite{schwenk2}, listed in Table \ref{AchimsLandaus}. These were calculated including the tensor interaction to second order.
We find that the exchange tensor, center of mass tensor and the cross vector terms are all quite small, so the matrix elements are only slightly modified from the value one obtains using only the central part of the interaction. In addition to being small, the tensor contributions have a different sign from the first order result:
\bea
\nonumber
\langle 01|A|21\rangle &\approx &0.07 \quad {\rm and}\\
\langle 21|A|21\rangle &\approx &1.06 \quad {\rm with}\nonumber\\
\langle 21|1+\mathcal{F}^{\rm c}|21\rangle &\approx &1.04 .
\eea

Since the calculation of tensor parameters in Ref. \cite{schwenk2} is only to second order, and the second order effects almost completely cancel the first order ones, higher order effects could be significant and should be calculated.
 
From the two sets of results above we can draw some conclusions regarding the effect of the tensor interaction on the susceptibility. The second term in Eq. (\ref{chiexpanded}) is clearly small compared to the first term, of the order of $1\%$, and may be neglected. The third and fourth terms, could be important, depending on the value of the tensor magnetic moment.  The corrections to the bare nucleon moment arise from two sources.  
The first is configuration mixing, which is due to the fact that a 
quasiparticle in the medium consists of a bare nucleon and a superposition of 
more complicated states involving a nucleon plus a number of particle-hole 
pairs. The second source is exchange currents, which are due to the fact that 
the external field can interact with other degrees of freedom than nucleons,
 for example, intermediate mesons in a nucleon-nucleon interaction.  For 
definiteness, we shall use the results given by Arima {\it et al}. in Ref. \cite{arima} for the 
Fermi gas model.  For neutrons in symmetric nuclear matter, they find 
$\mu\approx -1.62\ \mu_{\rm N}$ and $\mu_{\rm T} \approx 0.08\ \mu_{\rm N}$, where $\mu_{\rm N}=e\hbar/2M$ is the nuclear magneton.
Calculations of the configuration mixing contribution to weak interaction 
matrix elements made by Cowell and Pandharipande \cite{cowell} for 
asymmetric nuclear matter indicate that the renormalization of the magnetic 
moment does not depend strongly on proton fraction, and amounts to a 
suppression by about 10\%.  We would thus suspect that renormalization of 
the magnetic moment would be relatively insensitive to the proton fraction, 
and would be of a similar order of magnitude.  However, we stress that it is 
important to make more detailed estimates of magnetic moments in the nuclear 
medium.

The small magnitude of the tensor magnetic moment means that we may neglect 
also the third and fourth term in Eq. (\ref{chiexpanded}), both being less than one per cent of the first term. Therefore, we conclude that, the most important effects of the tensor force on the susceptibility is most likely the renormalization of the isotropic part of the magnetic moment and to the presence of transitions to multipair states. 

No reliable calculation of $\chi_{\rm M}$ exists, but by using the sum rule argument presented in Ref. \cite{olsson} we can put a lower bound on the 
multipair contribution to the susceptibility:
\beq
\chi_{\rm M}(q,0)\geq \frac{2 n S(q)}{\bar{\omega}},\qquad (q\rightarrow 0).
\label{Eq:multibound}
\eeq
Here $S(q)$ is the static structure function, $n$ is the density and $\bar{\omega}$ is the mean excitation energy. In Ref. \cite{olsson} we took the values of $S(q)=0.19$ and $\bar{\omega}= 63 \ {\rm MeV}$ from the calculations of Ref. \cite{akmal} and estimated the multipair contribution to the susceptibility to be more than $60 \%$ of the  susceptibility calculated by Fantoni et al \cite{Fantoni} ($\chi=0.38\chi_{\rm F}$). Here $\chi_{\rm F}$ is the susceptibility of a free Fermi gas of neutrons, $\chi_{\rm F}=3 n/2\epsilon_{\rm F}$. New calculations of the static structure function and the mean excitation energy by Cowell and Pandharipande \cite{cowell2} give a similar result, $S(q=0)=0.187$ and $\bar{\omega}=64\ {\rm MeV}$.
In order to compare with symmetric nuclear matter in next section we calculate $\chi_{\rm M}/\chi_{\rm F}$, and we find it to be $\sim 0.23$ for neutron matter, if we use the calculations of Ref. \cite{cowell2}. 

Finally, we note that the calculation for the total susceptibility of Fantoni et al. \cite{Fantoni} gives a result very close to the value one get by calculating $G_0$ from Brueckner theory and using Eq.\ (\ref{Landauchi}) for the susceptibility, which does not take into account the renormalization of the magnetic moment. This would appear to indicate that the tensor correlations redistribute spectral weight between single pair and multipair excitations, leaving the total susceptibility unchanged. A similar effect is familiar for the compressibility of electrons in metals when one takes into account the electron-phonon interaction \cite{leggett65b}.

\section{Symmetric nuclear matter}
\label{Sec:SNM}
Symmetric nuclear matter can be treated by a straightforward generalization of the discussion above 
by introducing the isospin degree of freedom in addition to the spin one.  Thus the generalization of  Eq. (\ref{landauint}) becomes
\bea
\nonumber
f_{\bf p p' \ve{\scriptstyle\sigma} \ve{\scriptstyle\sigma}'}&=& f_{\bf p p'} +
f^{'}_{\bf p p'}\ve{\tau}\cdot\ve{\tau}'+g_{\bf p p'}
\ve{\sigma}\cdot\ve{\sigma}'+g^{'}_{\bf p p'}
\ve{\sigma}\cdot\ve{\sigma}'\ve{\tau}\cdot\ve{\tau}'\\
&+&f^{\rm T}_{\bf p p' \ve{\scriptstyle\sigma}\ve{\scriptstyle \sigma}'}+f^{'\rm T}_{\bf p p' \ve{\scriptstyle\sigma}
\ve{\scriptstyle\sigma}'}\ve{\tau}\cdot\ve{\tau}',
\eea
where $\ve{\tau}$ is the isospin operator.  Since the magnetic moments of neutrons and protons are different, a magnetic field couples to both isoscalar and isovector components.  Because the difference between the magnetic moments of a neutron and a proton is much larger than the sum, the magnetic response will be dominated by the isovector term.  The isoscalar spin response is given by Eq. (\ref{chi}), but the density of states to be used is 
$N(0)=2m^*p_{\rm F}/\pi^2\hbar^3$, the factor of two reflecting the two nucleon species.  The isovector spin response is given by a similar expression, but with $G, H, K$ and $L$ replaced by  $G', H', K'$ and $L'$. Since there is no calculation of the new tensor terms for symmetric nuclear matter, we estimate the matrix elements by using the exchange tensor Landau parameters for symmetric nuclear matter calculated in Ref. \cite{baeckman}, which are listed in Table \ref{Tab:SNM}.
This gives the contribution to the matrix elements to be
\begin{table}
\caption{ Landau parameters for symmetric nuclear matter calculated by \cite{baeckman}\label{Tab:SNM}. The Landau parameters are calculated for Fermi wave vector $k_{\rm F}=1.35\ \rm{F}^{-1}$ and the effective mass is taken to be $m^*/m =1$.}
\begin{ruledtabular}
\begin{tabular}{lllll}
l & $G_l$ & $G_l'$ & $\tilde{H}_l$ & $\tilde{H}_l'$\\
\hline
0&0.447&1.291&0.65&-0.255\\
1&0.760&0.070&0.975&-0.359\\
2&0.276&0.090&0.829&-0.315\\
3&0.078&0.069&0.617&-0.233\\
\end{tabular}
\end{ruledtabular}
\end{table} 
\bea
\langle 01|A'|21\rangle&\approx& 0.11\nonumber \;\;\; {\rm and}\\
\langle 21|A'|21\rangle&\approx& 1.08.
\eea

The calculations of  Ref. \cite{arima} give for the  anisotropic contribution to the isovector magnetic moment 
the value $\mu_{\rm T}\approx 0.08\ \mu_{N}$ and therefore the third and fourth terms in the analogue 
of Eq. (\ref{chiexpanded}) for the isovector magnetic response may be neglected. The second term 
is also unimportant, since it is on the order of 1\% of the first term. We now use sum-rule arguments \cite{olsson} to put a bound on
the fifth term, which comes from multipair excitations.   
Equation (\ref{Eq:multibound}) applies to a Fermi system with an arbitary number of components, and recent calculations by Cowell and Pandharipande \cite{cowell2} of 
the static structure function and mean excitation energy for the isovector spin response. 
Their calculations give $S(q)=0.155$ and $\bar{\omega}\approx 253\ \rm{MeV}$ in the limit 
$q\rightarrow 0$. Comparing the contributions from multipair excitations with the susceptibility 
of a free Fermi gas we find
\beq
\frac{\chi_{\rm M}(q,0)}{\chi_{\rm F}}\geq \frac{4\epsilon_{\rm F} S(q)}{3 \bar{\omega}},
\eeq 
where $\chi_{\rm F}=3 n/2\epsilon_{\rm F}$ is the susceptibility of a free Fermi gas consisting of 2 species of spin-$1/2$ particles 
and $\epsilon_{\rm F}$ is the Fermi energy. At nuclear matter density $n=0.16\ \rm{fm}^{-3}$, $\epsilon_{\rm F} \approx 37 \ \rm{MeV}$, gives that $\chi_{\rm M}$ is at least $\sim 3 \ \%$ of $\chi_{\rm F}$, which is much smaller than the corresponding bound for pure neutrons.

We are not aware of recent calculations of the total susceptibility for symmetric nuclear matter similar to the calculations of Ref. \cite{Fantoni}, but we note that for neutron matter those calculations gave a susceptibility very close to the result one obtains by calculating $G_0$ from Brueckner theory and using Eq. (\ref{Landauchi}) to calculate the susceptibility. Therefore we compare $\chi_{\rm M}$ with the analogue for symmetric nuclear matter of Eq. (\ref{Landauchi}), which we call $\chi_0$, and find
\beq
\frac{\chi_{\rm M}}{\chi_0}\geq \frac{4\epsilon_{\rm F} S(q)}{3 \bar{\omega}}\frac{(1+G^{'}_{0})}{m^*/m}\approx 0.09,
\eeq
where we have taken $m^*/m=0.8$, and $G_0'$ from Ref. \cite{baeckman}, as listed in Table \ref{Tab:SNM}. This result indicates that the multipair contributions, being at least $9\ \%$ of the total susceptibility, can have some importance for symmetric nuclear matter, even though the bound is much smaller than the corresponding bound for neutron matter. 
It would be valuable to have explicit calculations of the multipair contributions to the susceptibility since it is unclear to what extent the bound provides a realistic estimate of the value of the quantity.

\section{Concluding remarks}
\label{Sec:conclusions}
In this paper we have derived an expression for the magnetic susceptibility of a Fermi liquid for a general interaction which conserves the total angular momentum and the total spin of the quasiparticle-quasihole pair. Apart from the contribution from single quasiparticle-quasihole pairs, which may be calculated using Landau theory, the susceptibility also contains a contributions from excitation of multipair states.  In addition to the exchange tensor term usually included, we have taken into account the new tensor terms found in Ref. \cite{schwenk2}.  We have also introduced an alternative parametrization of the exchange tensor interaction which has the advantage of reducing the importance of high angular harmonic contributions.

For neutron matter we find that the most important effects of the tensor force are to renormalize the magnetic moment and to give a contribution from multipair states. For symmetric nuclear matter one can draw the same conclusion, the bound on the multipair contribution is, however, much smaller, but it is not clear how close the actual contribution to the susceptibility, from multipairs, is to this lower bound.

The formalism may easily be extended to calculate the response to weak probes: the only difference being that the magnetic moments must be replaced by the corresponding matrix elements of the weak charges.

\begin{acknowledgments} We thank Achim Schwenk for interesting discussions and for providing us with his calculations of the tensor Landau parameters and Shannon Cowell and Vijay Pandharipande for giving us the result of their calculations of the static structure factor and energy-weighted sum. We are also grateful to Bengt Friman and Andrew Jackson for useful discussions and to Dan-Olof Riska for helpful correspondence. In addition, we thank ECT* and its director Wolfram Weise for  hospitality in Trento. 
One of the authors (PH) was partially supported by the KBN grant no. 2P03D.020.20, and of us (EO) acknowledges financial support in part from a European Commission Marie Curie Training Site Fellowship under Contract No. HPMT-2000-00100.  This work has been conducted within the framework of the school on
Advanced Instrumentation and Measurements (AIM) at Uppsala University
supported financially by the Foundation for Strategic Research (SSF).
\end{acknowledgments}

\bibliography{refs.bib}

\begin{thebibliography}{34}
\expandafter\ifx\csname natexlab\endcsname\relax\def\natexlab#1{#1}\fi
\expandafter\ifx\csname bibnamefont\endcsname\relax
  \def\bibnamefont#1{#1}\fi
\expandafter\ifx\csname bibfnamefont\endcsname\relax
  \def\bibfnamefont#1{#1}\fi
\expandafter\ifx\csname citenamefont\endcsname\relax
  \def\citenamefont#1{#1}\fi
\expandafter\ifx\csname url\endcsname\relax
  \def\url#1{\texttt{#1}}\fi
\expandafter\ifx\csname urlprefix\endcsname\relax\def\urlprefix{URL }\fi
\providecommand{\bibinfo}[2]{#2}
\providecommand{\eprint}[2][]{\url{#2}}

\bibitem[{\citenamefont{{Migdal}}(1968)}]{migdal}
\bibinfo{author}{\bibfnamefont{A.~B.} \bibnamefont{{Migdal}}},
  \emph{\bibinfo{title}{Nuclear Theory: The Quasiparticle Method}}
  (\bibinfo{publisher}{Benjamin, N.\ Y.}, \bibinfo{year}{1968}).

\bibitem[{\citenamefont{{D\c{a}browski} and {Haensel}}(1976)}]{dabrowski2}
\bibinfo{author}{\bibfnamefont{J.}~\bibnamefont{{D\c{a}browski}}}
  \bibnamefont{and}
  \bibinfo{author}{\bibfnamefont{P.}~\bibnamefont{{Haensel}}},
  \bibinfo{journal}{Ann. Phys.} \textbf{\bibinfo{volume}{97}},
  \bibinfo{pages}{452} (\bibinfo{year}{1976}).

\bibitem[{\citenamefont{{D\c{a}browski} and {Haensel}}(1974)}]{dabrowski1}
\bibinfo{author}{\bibfnamefont{J.}~\bibnamefont{{D\c{a}browski}}}
  \bibnamefont{and}
  \bibinfo{author}{\bibfnamefont{P.}~\bibnamefont{{Haensel}}},
  \bibinfo{journal}{Can. J. Phys} \textbf{\bibinfo{volume}{52}},
  \bibinfo{pages}{1768} (\bibinfo{year}{1974}).

\bibitem[{\citenamefont{{Haensel} and {D\c{a}browski}}(1975)}]{haensel2}
\bibinfo{author}{\bibfnamefont{P.}~\bibnamefont{{Haensel}}} \bibnamefont{and}
  \bibinfo{author}{\bibfnamefont{J.}~\bibnamefont{{D\c{a}browski}}},
  \bibinfo{journal}{Nucl. Phys. A} \textbf{\bibinfo{volume}{254}},
  \bibinfo{pages}{211} (\bibinfo{year}{1975}).

\bibitem[{\citenamefont{{B\"ackman} et~al.}(1979)\citenamefont{{B\"ackman},
  {Sj\"oberg}, and {Jackson}}}]{baeckman}
\bibinfo{author}{\bibfnamefont{S.-O.} \bibnamefont{{B\"ackman}}},
  \bibinfo{author}{\bibfnamefont{O.}~\bibnamefont{{Sj\"oberg}}},
  \bibnamefont{and} \bibinfo{author}{\bibfnamefont{A.~D.}
  \bibnamefont{{Jackson}}}, \bibinfo{journal}{Nucl. Phys. A}
  \textbf{\bibinfo{volume}{321}}, \bibinfo{pages}{10} (\bibinfo{year}{1979}).

\bibitem[{\citenamefont{{Schwenk} et~al.}(2002)\citenamefont{{Schwenk},
  {Brown}, and {Friman}}}]{schwenk}
\bibinfo{author}{\bibfnamefont{A.}~\bibnamefont{{Schwenk}}},
  \bibinfo{author}{\bibfnamefont{G.~E.} \bibnamefont{{Brown}}},
  \bibnamefont{and} \bibinfo{author}{\bibfnamefont{B.}~\bibnamefont{{Friman}}},
  \bibinfo{journal}{Nucl. Phys. A} \textbf{\bibinfo{volume}{703}},
  \bibinfo{pages}{745} (\bibinfo{year}{2002}).

\bibitem[{\citenamefont{{Schwenk} and {Friman}}(2004)}]{schwenk2}
\bibinfo{author}{\bibfnamefont{A.}~\bibnamefont{{Schwenk}}} \bibnamefont{and}
  \bibinfo{author}{\bibfnamefont{B.}~\bibnamefont{{Friman}}},
  \bibinfo{journal}{Phys. Rev. Lett.} \textbf{\bibinfo{volume}{92}},
  \bibinfo{pages}{082501} (\bibinfo{year}{2004}).

\bibitem[{\citenamefont{{Dickhoff} et~al.}(1983)\citenamefont{{Dickhoff},
  {Faessler}, {M\"unther}, and {Wu}}}]{dickhoff}
\bibinfo{author}{\bibfnamefont{W.~H.} \bibnamefont{{Dickhoff}}},
  \bibinfo{author}{\bibfnamefont{A.}~\bibnamefont{{Faessler}}},
  \bibinfo{author}{\bibfnamefont{H.}~\bibnamefont{{M\"unther}}},
  \bibnamefont{and} \bibinfo{author}{\bibfnamefont{S.-S.} \bibnamefont{{Wu}}},
  \bibinfo{journal}{Nucl. Phys. A} \textbf{\bibinfo{volume}{405}},
  \bibinfo{pages}{534} (\bibinfo{year}{1983}).

\bibitem[{\citenamefont{{Janka}}(2001)}]{janka}
\bibinfo{author}{\bibfnamefont{H.-T.} \bibnamefont{{Janka}}},
  \bibinfo{journal}{Astrophys. J.} \textbf{\bibinfo{volume}{368}},
  \bibinfo{pages}{527} (\bibinfo{year}{2001}).

\bibitem[{\citenamefont{{Sawyer}}(1975)}]{sawyer}
\bibinfo{author}{\bibfnamefont{R.~F.} \bibnamefont{{Sawyer}}},
  \bibinfo{journal}{Phys. Rev. D} \textbf{\bibinfo{volume}{11}},
  \bibinfo{pages}{2740} (\bibinfo{year}{1975}).

\bibitem[{\citenamefont{{Iwamoto} and {Pethick}}(1982)}]{iwamoto}
\bibinfo{author}{\bibfnamefont{N.}~\bibnamefont{{Iwamoto}}} \bibnamefont{and}
  \bibinfo{author}{\bibfnamefont{C.~J.} \bibnamefont{{Pethick}}},
  \bibinfo{journal}{Phys. Rev. D} \textbf{\bibinfo{volume}{25}},
  \bibinfo{pages}{313} (\bibinfo{year}{1982}).

\bibitem[{\citenamefont{{Raffelt} and {Seckel}}(1995)}]{raffelt1}
\bibinfo{author}{\bibfnamefont{G.}~\bibnamefont{{Raffelt}}} \bibnamefont{and}
  \bibinfo{author}{\bibfnamefont{D.}~\bibnamefont{{Seckel}}},
  \bibinfo{journal}{Phys. Rev. D} \textbf{\bibinfo{volume}{52}},
  \bibinfo{pages}{1780} (\bibinfo{year}{1995}).

\bibitem[{\citenamefont{{Raffelt} et~al.}(1996)\citenamefont{{Raffelt},
  {Seckel}, and {Sigl}}}]{raffelt2}
\bibinfo{author}{\bibfnamefont{G.}~\bibnamefont{{Raffelt}}},
  \bibinfo{author}{\bibfnamefont{D.}~\bibnamefont{{Seckel}}}, \bibnamefont{and}
  \bibinfo{author}{\bibfnamefont{G.}~\bibnamefont{{Sigl}}},
  \bibinfo{journal}{Phys. Rev. D} \textbf{\bibinfo{volume}{54}},
  \bibinfo{pages}{2784} (\bibinfo{year}{1996}).

\bibitem[{\citenamefont{{Raffelt} and {Sigl}}(1999)}]{raffelt3}
\bibinfo{author}{\bibfnamefont{G.}~\bibnamefont{{Raffelt}}} \bibnamefont{and}
  \bibinfo{author}{\bibfnamefont{G.}~\bibnamefont{{Sigl}}},
  \bibinfo{journal}{Phys. Rev. D} \textbf{\bibinfo{volume}{60}},
  \bibinfo{pages}{023001} (\bibinfo{year}{1999}).

\bibitem[{\citenamefont{{Burrows} and {Sawyer}}(1998)}]{burrows}
\bibinfo{author}{\bibfnamefont{A.}~\bibnamefont{{Burrows}}} \bibnamefont{and}
  \bibinfo{author}{\bibfnamefont{R.~F.} \bibnamefont{{Sawyer}}},
  \bibinfo{journal}{Phys. Rev. C} \textbf{\bibinfo{volume}{58}},
  \bibinfo{pages}{554} (\bibinfo{year}{1998}).

\bibitem[{\citenamefont{{Prakash} et~al.}(2001)\citenamefont{{Prakash},
  {Lattimer}, {Sawyer}, and {Volkas}}}]{prakash}
\bibinfo{author}{\bibfnamefont{M.}~\bibnamefont{{Prakash}}},
  \bibinfo{author}{\bibfnamefont{J.~M.} \bibnamefont{{Lattimer}}},
  \bibinfo{author}{\bibfnamefont{R.~F.} \bibnamefont{{Sawyer}}},
  \bibnamefont{and} \bibinfo{author}{\bibfnamefont{R.~R.}
  \bibnamefont{{Volkas}}}, \bibinfo{journal}{Annu. Rev. Nucl. Part. Sci.}
  \textbf{\bibinfo{volume}{51}}, \bibinfo{pages}{295} (\bibinfo{year}{2001}).

\bibitem[{\citenamefont{{Reddy} et~al.}(1998)\citenamefont{{Reddy}, {Prakash},
  and {Lattimer}}}]{prakashreddy}
\bibinfo{author}{\bibfnamefont{S.}~\bibnamefont{{Reddy}}},
  \bibinfo{author}{\bibfnamefont{M.}~\bibnamefont{{Prakash}}},
  \bibnamefont{and} \bibinfo{author}{\bibfnamefont{J.~M.}
  \bibnamefont{{Lattimer}}}, \bibinfo{journal}{Phys. Rev. D}
  \textbf{\bibinfo{volume}{58}}, \bibinfo{pages}{013009}
  (\bibinfo{year}{1998}).

\bibitem[{\citenamefont{{Fantoni} et~al.}(2001)\citenamefont{{Fantoni},
  {Sarsa}, and {Schmidt}}}]{Fantoni}
\bibinfo{author}{\bibfnamefont{S.}~\bibnamefont{{Fantoni}}},
  \bibinfo{author}{\bibfnamefont{A.}~\bibnamefont{{Sarsa}}}, \bibnamefont{and}
  \bibinfo{author}{\bibfnamefont{K.~E.} \bibnamefont{{Schmidt}}},
  \bibinfo{journal}{Phys. Rev. Lett.} \textbf{\bibinfo{volume}{87}},
  \bibinfo{pages}{181101} (\bibinfo{year}{2001}).

\bibitem[{\citenamefont{{Olsson} and {Pethick}}(2002)}]{olsson}
\bibinfo{author}{\bibfnamefont{E.}~\bibnamefont{{Olsson}}} \bibnamefont{and}
  \bibinfo{author}{\bibfnamefont{C.~J.} \bibnamefont{{Pethick}}},
  \bibinfo{journal}{Phys. Rev. C} \textbf{\bibinfo{volume}{66}},
  \bibinfo{pages}{065803} (\bibinfo{year}{2002}).

\bibitem[{\citenamefont{{Miyazawa}}(1951)}]{miyazawa}
\bibinfo{author}{\bibfnamefont{H.}~\bibnamefont{{Miyazawa}}},
  \bibinfo{journal}{Prog. Theor. Phys.} \textbf{\bibinfo{volume}{6}},
  \bibinfo{pages}{801} (\bibinfo{year}{1951}).

\bibitem[{\citenamefont{{Arima} et~al.}(1987)\citenamefont{{Arima}, {Shimizu},
  {Bentz}, and {Hyuga}}}]{arima}
\bibinfo{author}{\bibfnamefont{A.}~\bibnamefont{{Arima}}},
  \bibinfo{author}{\bibfnamefont{K.}~\bibnamefont{{Shimizu}}},
  \bibinfo{author}{\bibfnamefont{W.}~\bibnamefont{{Bentz}}}, \bibnamefont{and}
  \bibinfo{author}{\bibfnamefont{H.}~\bibnamefont{{Hyuga}}},
  \bibinfo{journal}{Adv. Nucl. Phys.} \textbf{\bibinfo{volume}{18}},
  \bibinfo{pages}{1} (\bibinfo{year}{1987}).

\bibitem[{\citenamefont{{Tsushima} et~al.}(1993)\citenamefont{{Tsushima},
  {Riska}, and {Blunden}}}]{riska}
\bibinfo{author}{\bibfnamefont{K.}~\bibnamefont{{Tsushima}}},
  \bibinfo{author}{\bibfnamefont{D.~O.} \bibnamefont{{Riska}}},
  \bibnamefont{and} \bibinfo{author}{\bibfnamefont{P.~G.}
  \bibnamefont{{Blunden}}}, \bibinfo{journal}{Nucl. Phys. A}
  \textbf{\bibinfo{volume}{559}}, \bibinfo{pages}{543} (\bibinfo{year}{1993}).

\bibitem[{\citenamefont{{Haensel} and {Jerzak}}(1982)}]{haensel}
\bibinfo{author}{\bibfnamefont{P.}~\bibnamefont{{Haensel}}} \bibnamefont{and}
  \bibinfo{author}{\bibfnamefont{A.~J.} \bibnamefont{{Jerzak}}},
  \bibinfo{journal}{Phys. Lett. B} \textbf{\bibinfo{volume}{112}},
  \bibinfo{pages}{285} (\bibinfo{year}{1982}).

\bibitem[{\citenamefont{{Leggett}}(1965)}]{leggett65b}
\bibinfo{author}{\bibfnamefont{A.~J.} \bibnamefont{{Leggett}}},
  \bibinfo{journal}{Phys. Rev.} \textbf{\bibinfo{volume}{140}},
  \bibinfo{pages}{1869} (\bibinfo{year}{1965}).

\bibitem[{\citenamefont{{Baym} and {Pethick}}(1991)}]{baym}
\bibinfo{author}{\bibfnamefont{G.}~\bibnamefont{{Baym}}} \bibnamefont{and}
  \bibinfo{author}{\bibfnamefont{C.~J.} \bibnamefont{{Pethick}}},
  \emph{\bibinfo{title}{Landau Fermi-liquid theory: concepts and applications}}
  (\bibinfo{publisher}{Wiley N. Y.}, \bibinfo{year}{1991}),
  p.~\bibinfo{pages}{71}.

\bibitem[{\citenamefont{{Landau}}(1957{\natexlab{a}})}]{landau1}
\bibinfo{author}{\bibfnamefont{L.~D.} \bibnamefont{{Landau}}},
  \bibinfo{journal}{Sov. Phys. JETP} \textbf{\bibinfo{volume}{3}},
  \bibinfo{pages}{920} (\bibinfo{year}{1957}{\natexlab{a}}).

\bibitem[{\citenamefont{{Landau}}(1957{\natexlab{b}})}]{landau2}
\bibinfo{author}{\bibfnamefont{L.~D.} \bibnamefont{{Landau}}},
  \bibinfo{journal}{Sov. Phys. JETP} \textbf{\bibinfo{volume}{5}},
  \bibinfo{pages}{101} (\bibinfo{year}{1957}{\natexlab{b}}).

\bibitem[{\citenamefont{{Pethick} and {Carneiro}}(1973)}]{carneiro1}
\bibinfo{author}{\bibfnamefont{C.~J.} \bibnamefont{{Pethick}}}
  \bibnamefont{and} \bibinfo{author}{\bibfnamefont{G.~M.}
  \bibnamefont{{Carneiro}}}, \bibinfo{journal}{Phys. Rev. A}
  \textbf{\bibinfo{volume}{7}}, \bibinfo{pages}{304} (\bibinfo{year}{1973}).

\bibitem[{\citenamefont{{Pethick} and {Carneiro}}(1975)}]{carneiro2}
\bibinfo{author}{\bibfnamefont{C.~J.} \bibnamefont{{Pethick}}}
  \bibnamefont{and} \bibinfo{author}{\bibfnamefont{G.~M.}
  \bibnamefont{{Carneiro}}}, \bibinfo{journal}{Phys. Rev. B}
  \textbf{\bibinfo{volume}{11}}, \bibinfo{pages}{1106} (\bibinfo{year}{1975}).

\bibitem[{\citenamefont{{Schwenk}}(2003)}]{schwenk3}
\bibinfo{author}{\bibfnamefont{A.}~\bibnamefont{{Schwenk}}}
  (\bibinfo{year}{2003}), \bibinfo{note}{private communication}.

\bibitem[{\citenamefont{{Schwenk} et~al.}(2003)\citenamefont{{Schwenk},
  {Friman}, and {Brown}}}]{schwenk1}
\bibinfo{author}{\bibfnamefont{A.}~\bibnamefont{{Schwenk}}},
  \bibinfo{author}{\bibfnamefont{B.}~\bibnamefont{{Friman}}}, \bibnamefont{and}
  \bibinfo{author}{\bibfnamefont{G.~E.} \bibnamefont{{Brown}}},
  \bibinfo{journal}{Nucl. Phys. A} \textbf{\bibinfo{volume}{713}},
  \bibinfo{pages}{191} (\bibinfo{year}{2003}).

\bibitem[{\citenamefont{{Cowell} and {Pandharipande}}(2003)}]{cowell}
\bibinfo{author}{\bibfnamefont{S.}~\bibnamefont{{Cowell}}} \bibnamefont{and}
  \bibinfo{author}{\bibfnamefont{V.~R.} \bibnamefont{{Pandharipande}}},
  \bibinfo{journal}{Phys. Rev. C} \textbf{\bibinfo{volume}{67}},
  \bibinfo{pages}{035504} (\bibinfo{year}{2003}).

\bibitem[{\citenamefont{{Akmal} and {Pandharipande}}(1997)}]{akmal}
\bibinfo{author}{\bibfnamefont{A.}~\bibnamefont{{Akmal}}} \bibnamefont{and}
  \bibinfo{author}{\bibfnamefont{V.~R.} \bibnamefont{{Pandharipande}}},
  \bibinfo{journal}{Phys. Rev. C} \textbf{\bibinfo{volume}{56}},
  \bibinfo{pages}{2261} (\bibinfo{year}{1997}).

\bibitem[{\citenamefont{{Cowell} and {Pandharipande}}(2004)}]{cowell2}
\bibinfo{author}{\bibfnamefont{S.}~\bibnamefont{{Cowell}}} \bibnamefont{and}
  \bibinfo{author}{\bibfnamefont{V.~R.} \bibnamefont{{Pandharipande}}}
  (\bibinfo{year}{2004}), \bibinfo{note}{private communication}.

\end{thebibliography}

\end{document}